# Magnetic field stabilized Wigner crystal states in a graphene moiré superlattice

Guorui Chen[1,2*], Ya-Hui Zhang[3], Aaron Sharpe[4,5,6], Zuocheng Zhang[2], Shaoxin Wang[2], Lili Jiang[2], Bosai Lyu[1], Hongyuan Li[2], Kenji Watanabe[7], Takashi Taniguchi[8], Zhiwen Shi[1], David Goldhaber-Gordon[5,9], Yuanbo Zhang[10,11*], Feng Wang[2,12,13*]

[1]*Key Laboratory of Artificial Structures and Quantum Control (Ministry of Education), Shenyang National Laboratory for Materials Science, School of Physics and Astronomy, Shanghai Jiao Tong University, Shanghai 200240, China.*

[2]*Department of Physics, University of California at Berkeley, Berkeley, CA 94720, USA.*

[3]*Department of Physics, Harvard University, Cambridge, MA 02138, USA.*

[4]*Department of Applied Physics, Stanford University, Stanford, CA 94305, USA.*

[5]*Stanford Institute for Materials and Energy Sciences, SLAC National Accelerator Laboratory, Menlo Park, CA 94025, USA.*

[6]*Quantum and Electronic Materials Department, Sandia National Laboratories, Livermore CA 94550, USA*

[7]*Research Center for Functional Materials, National Institute for Materials Science, 1-1 Namiki, Tsukuba 305-0044, Japan.*

[8]*International Center for Materials Nanoarchitectonics, National Institute for Materials Science, 1-1 Namiki, Tsukuba 305-0044, Japan.*

[9]*Department of Physics, Stanford University, Stanford, CA 94305, USA.*

[10]*State Key Laboratory of Surface Physics and Department of Physics, Fudan University, Shanghai 200433, China.*

[11]*Institute for Nanoelectronic Devices and Quantum Computing, Fudan University, Shanghai 200433, China.*

[12]*Materials Science Division, Lawrence Berkeley National Laboratory, Berkeley, CA 94720, USA.*

[13]*Kavli Energy NanoSciences Institute at the University of California, Berkeley and the Lawrence Berkeley National Laboratory, Berkeley, CA 94720, USA.*

*Correspondence to: chenguorui@sjtu.edu.cn, zhyb@fudan.edu.cn, fengwang76@berkeley.edu

**Wigner crystals are predicted as the crystallization of the dilute electron gas moving in a uniform background when the electron-electron Coulomb energy dominates the kinetic energy**[1]. **The Wigner crystal has previously been observed in the ultraclean two-dimensional electron gas (2DEG) present on the surface of liquid helium**[2,3] **and in semiconductor quantum wells at high magnetic field**[4,5]. **More recently, Wigner crystals have also been reported in $WS_2/WSe_2$ moiré heterostructures**[6–8]. **ABC-stacked trilayer graphene on boron nitride (ABC-TLG/hBN) moiré superlattices provide a unique tunable platform to explore Wigner crystal states where the electron correlation can be controlled by electric and magnetic field. Here we report the observation of magnetic field stabilized Wigner crystal states in a ABC-TLG/hBN moiré superlattice. We show that correlated insulating states emerge at multiple fractional and integer fillings corresponding to $\nu$ = 1/3, 2/3, 1, 4/3, 5/3 and 2 electrons per moiré lattice site under a magnetic field. These correlated insulating states can be attributed to generalized Mott states for the integer fillings ($\nu$ = 1, 2) and generalized Wigner crystal states for the fractional fillings ($\nu$ = 1/3, 2/3, 4/3, 5/3). The generalized Wigner crystal states are stabilized by a vertical magnetic field, and they are strongest at one magnetic flux quantum per three moiré superlattices. The correlated insulating states at $\nu$ = 2 persists up to 30 Tesla, which can be described by a Mott-Hofstadter transition at high magnetic field. The tunable Mott and Wigner crystal states in the ABC-TLG/hBN highlight the opportunities to discover new correlated quantum phases due to the interplay between the magnetic field and moiré flatbands.**

Moiré superlattices in van der Waals heterostructures are emerging as a powerful platform for exploring novel correlated electron physics, where the moiré bandwidths can be made small enough that electron correlations dominate[9,10]. The ABC-stacked trilayer graphene on hBN (ABC-TLG/hBN) moiré heterostructure provides a remarkably tunable system to explore the interplay between electron-electron interactions, the moiré flatband bandwidth, and the moiré band topology, where Mott insulator, superconductor, Chern insulator, and fractional-filling orbital ferromagnet phases can all be realized in a single device through electrical control[11–14]. The magnetic field provides another dimension to manipulate correlated quantum phases in moiré heterostructures. Fascinating Hofstadter butterfly states have been previously observed in highly dispersive monolayer graphene/hBN moiré superlattices in high magnetic field when the cyclotron orbit size is comparable to the moiré period[15–18]. In ABC-TLG/hBN moiré superlattice, we have the combination of a magnetic field with a topological and flat moiré minibands[19,20]. It can lead to new tunable quantum phases distinctly different from the conventional Hofstadter butterfly phenomena.

Here we report the observation of generalized Wigner crystal states stabilized by the magnetic field at fractional electron fillings in an ABC-TLG/hBN moiré superlattice. Upon applying a perpendicular magnetic field $(B\perp)$, new insulating states emerge at the electron doping of $n = n_0/3$, $2n_0/3$, $n_0$, $4n_0/3$, $5n_0/3$ and $2n_0$ in the ABC-TLG/hBN moiré superlattice, where $n_0$ corresponds to one electron per moiré site. These insulating states can be understood as generalized Mott states at integer fillings ($n = n_0$ and $2n_0$) and generalized Wigner crystal states at fractional fillings ($n = n_0/3$, $2n_0/3$, $4n_0/3$ and $5n_0/3$). The generalized Wigner crystal states arise from an unusual interplay between the magnetic field and the moiré bandwidth: they require both a sufficiently flat moiré miniband at large vertical displacement field $D$ and a perpendicular $B\perp$ close to 6 Tesla. The Mott insulating state at $n = 2n_0$, however, does not have a strong $B\perp$ field dependence. It persists up to 30 Tesla, which can be understood as the Hofstadter butterfly state in the high magnetic field.

Figure 1a shows a schematic of our ABC-TLG/hBN heterostructure device, which is the same sample as in Ref 13. The ABC-TLG is aligned with the bottom hBN crystal with zero twisted angle. The moiré superlattice period is 15 nm. Gate voltages $V_t$ and $V_b$ separately applied to the metal top gate and the Si bottom gate allow us to independently control the doping $n$ and the miniband bandwidth tuned by the vertical electric field $D$ following the relations: $n = (D_b - D_t)/e$, and $D = (D_b + D_t)/2$.[21] Here $D_b = +\varepsilon_b(V_b - V_b^0)/d_b$ and $D_t = -\varepsilon_t(V_t - V_t^0)/d_t$ are the vertical displacement field below and above the ABC-TLG/hBN, respectively, $\varepsilon_{b(t)}$ and $d_{b(t)}$ are the dielectric constant and thickness of the bottom (top) dielectric layers, and $V_{b(t)}^0$ is the effective offset in the bottom (top) gate voltages caused by environment-induced carrier doping. Figure 1b shows the calculated single-particle bandstructure for ABC-TLG/hBN at $B\perp = 0$ and $D = 0.5$ V/nm.[13] The lowest valance miniband tend to have narrower bandwidth and therefore stronger correlation effects than the lowest conduction miniband at zero magnetic field.

Figure 1c shows the experimental data of the longitudinal resistivity $\rho_{xx}$ in the ABC-TLG/hBN moiré superlattice as a function of $n$ and $D$. Resistivity peaks at $n = 0$ and $n = \pm 4n_0$ correspond to charge neutrality point and fully filled points of conduction and valence single-particle minibands, respectively. Correlated insulator states can be clearly observed at hole doping of $n = -n_0$ and $n = -2n_0$ when the displacement field $|D|$ is larger than ~ 0.2 V/nm. For electron doping, the correlated insulator state appears at $n = 2n_0$ only when $D$ is more negative than -0.6 V/nm. The large electron-hole asymmetry can be accounted for by the flatter valence moiré miniband at zero magnetic field, as shown in Fig. 1b. In the same ABC-TLG/hBN moiré heterostructure, signatures of superconductivity has been observed near $n = -n_0$ at large positive $D$,[12] and a $C = 2$ Chern insulator has been observed at $n = -n_0$ with a topological flat moiré miniband at negative $D$.[13]

The correlated electron phenomena change dramatically under a perpendicular magnetic field $B\perp$. Figure 2a shows the magneto-transport data of $\rho_{xx}$ as functions of $n$ and $B\perp$ at $D = -0.4$V/nm where one expects the bands to be topologically non-trivial.

The correlated insulator states in the valence moiré miniband ($n < 0$) are destabilized for $B\perp > 4$ T. Simultaneously, electron correlation becomes stronger at finite $B\perp$ for the conduction moiré miniband ($n > 0$). First, two prominent $\rho_{xx}$ peaks appear at the integer fillings of $n = n_0$ and $2n_0$ for $B\perp > 3$ T, corresponding to one and two electrons per moiré site, respectively. Then, multiple resistivity peaks at fractional fillings of $n_0/3$, $2n_0/3$, $4n_0/3$ and $5n_0/3$ of the conduction minibands, corresponding to 1, 2, 4 and 5 electrons per *three* moiré sites, emerge at $B\perp$ between 5 to 8 Tesla.

These magnetic field stabilized insulating states exhibit several notable features. First, all the resistivity peaks occur at near constant fillings and their positions are largely independent on the magnetic field. This behavior is distinctly different from quantized Landau levels, where the carrier density in a Landau level scales linearly with $B\perp$.[22] Second, the $\rho_{xx}$ peaks at fillings from $n_0/3$ to $5n_0/3$ are maximum near $B\perp = 5.5$ T, as shown in Fig. 2b. This magnetic field corresponds to a magnetic flux $\Phi = B_\perp \cdot A = 0.78\Phi_0$ through three moiré lattice area, where $\Phi_0 = h/e$ is the magnetic flux quantum. Third, the resistivity peaks within the conduction band appear with the application of an out-of-plane magnetic field $B\perp$, but not an in-plane magnetic field $B_{//}$ (see Fig. S1 for $B_{//}$ data).

These magnetic-field stabilized insulating states observed at integer and fractional fillings are tunable by the displacement field $D$, which controls the bandwidth of the moiré minibands in ABC-TLG/hBN. Figure 2c shows the evolution of $\rho_{xx}$ as a function of the displacement field $D$ for a fixed $B\perp = 5.5$ T. The resistivity peaks first appear at $D \sim -0.2$ V/nm, and become increasingly prominent with increasing the amplitude of $D$ at fixed $B\perp = 5.5$ T. At the highest $D$ field, even more insulating states beyond the $n/3$ fillings start to emerge. Figure 2d shows normalized $\rho_{xx}$ as a function of $D$ at integer and fractional fillings. These $B\perp$ stabilized insulating states exhibit similar $\rho_{xx}$ dependence on the $D$ field compared with that of the Mott insulating states of the valence band at zero magnetic field[11]. It suggests that the $B\perp$ stabilized insulating states in the conduction moiré minibands are also correlated insulators controlled by dominating electron-electron interactions when the moiré miniband becomes sufficiently flat.

The $B\perp$ stabilized insulating states are only observed with a negative $D$, not positive $D$, in the ABC-TLG/hBN moiré device. (See Fig. S2 for the comparison of the magneto transport at positive and negative $D$.) The asymmetry of the $D$ dependence originates from the fact that the moiré superlattice only exists between ABC-TLG and the bottom hBN. Figure S3b and c display the calculated LL fan for electron and hole doping in our structure. The theory predicts that the crossing LLs due to Hofstadter butterfly only exists in the electron side for $D < 0$ and in hole side for $D > 0$.

Figure 3a plots the $n$ dependent $\rho_{xx}$ curves at different temperatures for fixed $D = $ -0.4 V/nm and $B\perp = 5.5$ T. The resistivity peaks show clear insulating behavior with increased resistivity at lower temperatures. The Arrhenius plots of the temperature dependent $\rho_{xx}$ are shown in the inset of Fig. 3a. These Arrhenius plots exhibit similar slopes, indicating comparable transport thermal excitation gaps ~ 6 Kelvin for these $B\perp$ stabilized insulators at both fractional and integer fillings. We note that this transport gap may not be intrinsic because the transport behavior could be affected by extrinsic effects such as impurity scatterings.

We attribute the observed correlated states at integer fillings of $n = n_0$ and $2n_0$ to generalized Mott states and fractional fillings of $n = n_0/3$, $2n_0/3$, $4n_0/3$, $5n_0/3$ to generalized Wigner crystal states stabilized by the magnetic field. Fig. 3b illustrates the real space electron lattice of the insulating states at the fillings of $n = n_0/3$, $2n_0/3$, $n_0$ of the ABC-TLG/hBN moiré superlattice. There is one electron per three moiré sites at $n = n_0/3$, where the electrons not only avoid double occupancy at the same site, but also avoid simultaneous occupation of the adjacent moiré sites. This state is known as a generalized Wigner crystal state. It highlights the importance of long-range electron interactions beyond the on-site Coulomb potential as typically used in the Hubbard model. The generalized Wigner crystal states at fractional fillings are most prominent when there is approximately one magnetic flux quantum per Wigner crystal lattice (i.e. three times of a moiré cell).

The integer filling state at $2n_0$ persists to very high magnetic field up to 30 T, as shown in Fig. S3a, which we attribute to the Hofstadter butterfly state. The magnetic field dependence of the $2n_0$ is very different to the other states (Fig. 2a, b and Fig. S3a), suggesting a different origin in the very high magnetic field. We calculate the Hofstadter butterfly states of the ABC-TLG/hBN with the potential difference of $\pm 20$ meV between the top and bottom graphene layers (corresponding to $D = \pm 0.4$ V/nm, see Methods for calculation details). Our calculation (within the single-particle picture) predicts that LL at $2n_0$ with a filling factor $\nu = 0$ is developed at $B\perp = 7$ T and persists in the high magnetic field up to 30 T (Fig. S3b), which accounts for the stability of the insulating state at $n = 2n_0$ in very high field. Therefore, the insulating state at $n = 2n_0$ is characterized by a correlated Mott insulator at low $B\perp$ and a Hofstadter butterfly insulator at high $B\perp$. In comparison, no LL and Hofstadter butterfly is predicted at $n = n_0$ in the single particle calculation (Fig. S3b). The insulating state at $n = n_0$ is a Mott state that is not related to the Hofstadter butterfly pattern.

Our observed Wigner crystal states are different from the quantum Hall Wigner crystals observed previously. In quantum Hall Wigner crystals, the carrier density of the Wigner crystal scales linearly with the magnetic field. Our generalized Wigner crystal states, however, occur at a constant carrier density as a function of field. The Wigner crystal lattice period of $\sqrt{3}L_M$ is locked to the moiré period and independent of the applied magnetic field. The locking between the Wigner crystal lattice and the moiré lattice has been observed recently in the TMD moiré heterostructures probed by optical measurements[6–8]. It is interesting to compare the Wigner crystal states in these two moiré systems. The electron lattices of Wigner crystals in both systems are triangular lattices with the periods of three times of the moiré periods. The strongest Wigner crystal states in these two systems are generally the same at $n_0/3$, $2n_0/3$, $4n_0/3$ and $5n_0/3$. However, there are some significant differences. The Wigner crystal states in TMD moiré superlattices originate from very strong electron-electron interactions. It does not depend sensitively to the external electrical and magnetic field. The Wigner crystal states in ABC-TLG/hBN moiré superlattice, however, are highly tunable by both the external magnetic field and the vertical electrical field. They offer an excellent platform to explore how different quantum phases compete and how the Wigner crystal states emerge at certain parameter ranges.

**Acknowledgments** G.C. acknowledges financial support from National Key Research Program of China (grant nos. 2020YFA0309000, 2021YFA1400100), NSF of China (grant no.12174248) and SJTU NO. 21X010200846. F.W. were supported as part of the Center for Novel Pathways to Quantum Coherence in Materials, an Energy Frontier Research Center funded by the US Department of Energy, Office of Science, Basic Energy Sciences. Z.C.Z. was supported by ARO grant W911NF2110176. A.L.S. was supported by a National Science Foundation Graduate Research Fellowship and a Ford Foundation Predoctoral Fellowship. The work of D.G.-G. on this project was supported by the US Department of Energy, Office of Science, Basic Energy Sciences, Materials Sciences and Engineering Division, under contract number DE-AC02-76SF00515. Y.Z. acknowledges financial support from National Key Research Program of China (grant nos. 2016YFA0300703, 2018YFA0305600), NSF of China (grant nos. U1732274, 11527805 and 11421404), Shanghai Municipal Science and Technology Commission (grant nos. 18JC1410300 and 2019SHZDZX01) and Strategic Priority Research Program of Chinese Academy of Sciences (grant no. XDB30000000). Z.S. acknowledges support from National Key Research and Development Program of China (grant number 2016YFA0302001) and National Natural Science Foundation of China (grant number 11774224, 12074244), and additional support from a Shanghai talent program. Growth of hexagonal boron nitride crystals was supported by the Elemental Strategy Initiative conducted by the MEXT, Japan, Grant Number JPMXP0112101001, JSPS KAKENHI Grant Number JP20H00354, 19H05790 and A3

Foresight by JSPS. Part of the sample fabrication was conducted at Nano-fabrication Laboratory at Fudan University. A portion of this work was performed at the National High Magnetic Field Laboratory, which is supported by the National Science Foundation Cooperative Agreement No. DMR-1644779 and the state of Florida. Sandia National Laboratories is a multi-mission laboratory managed and operated by National Technology & Engineering Solutions of Sandia, LLC, a wholly owned subsidiary of Honeywell International Inc., for the U.S. Department of Energy's National Nuclear Security Administration under contract DE-NA0003525. This paper describes objective technical results and analysis. Any subjective views or opinions that might be expressed in the paper do not necessarily represent the views of the U.S. Department of Energy or the United States Government.

**Author contributions**

G.C. and F.W. conceived the project. F.W. and Y.Z. supervised the project. G.C. fabricated samples and performed transport measurement with assistance from Z.Z., S.W. and A.S.. G.C., L.J., B.L., H.L. and Z.S. prepared trilayer graphene and performed near-field infrared and AFM measurements. K.W. and T.T. grew hBN single crystals. YH.Z. calculated the Landau diagram. G.C. and F.W. analyzed the data. G.C., YH.Z., and F.W. wrote the paper, with input from all authors.

**Author Information**

The authors declare no competing financial interests. Correspondence and requests for materials should be addressed to G.C. (chenguorui@sjtu.edu.cn), F.W. (fengwang76@berkeley.edu), and Y.Z. (zhyb@fudan.edu.cn).

**Methods**

**Transport measurements.** The electrical transport measurement below 14 Tesla is performed in Oxford Instruments superconducting magnet systems. Measurement above 14 Tesla is performed at Maglab in Tallahassee. Stanford Research Systems

SR830 lock-in amplifiers are used to measure the resistivity of the device with an AC bias current of 1 nA at a frequency of ~17 Hz.

**Calculation for Hofstadter butterfly states.** The total Hamiltonian

$$H = H_0 + H_M$$

For the purpose of Hofstadter bands, we use a simple $2 \times 2$ model for the ABC-TLG based on $A_1$, $B_3$ orbitals. An effective model for graphene at K valley is

$$H_0 = \left(f_{A_1}^\dagger(\mathbf{k}), f_{B_3}^\dagger(\mathbf{k})\right)$$

$$\begin{pmatrix} \Phi_V + \dfrac{2vv_4 k^2}{\gamma_1} & \dfrac{v^N}{(-\gamma_1)^{N-1}}(\pi^\dagger)^N + \gamma_2 + \dfrac{2vv_3 k^2}{\gamma_1} \\ \dfrac{v^N}{(-\gamma_1)^{N-1}}\pi^N + \gamma_2 + \dfrac{2vv_3 k^2}{\gamma_1} & -\Phi_V + \dfrac{2vv_4 k^2}{\gamma_1} \end{pmatrix} \begin{pmatrix} f_{A_1}(\mathbf{k}) \\ f_{B_3}(\mathbf{k}) \end{pmatrix}$$

We have $\pi = k_x + ik_y$, $v = \frac{\sqrt{3}}{2}|t|$, $v_3 = \frac{\sqrt{3}}{2}\gamma_3$, and $v_4 = \frac{\sqrt{3}}{2}\gamma_4$, where $(t, \gamma_1, \gamma_2, \gamma_3, \gamma_4) = $ (-2610, 358, -8.3, 293, 144) meV. The Hamiltonian at finite magnetic field $B$ can be obtained by substituting $\pi_x = p_x - eA_x$ and $\pi_y = p_y - eA_y$.

The moiré superlattice potential is

$$H_M = \sum_{j=1,2,3} \left(Ve^{i\varphi}\right) f_{A_1}^\dagger(\mathbf{k} + \mathbf{Q}_j) + h.c.$$

Where $\mathbf{Q}_1 = (0, \frac{4\pi}{\sqrt{3}a_M})$, $\mathbf{Q}_2 = (-\frac{2\pi}{a_M}, \frac{2\pi}{\sqrt{3}a_M})$, and $\mathbf{Q}_3 = (-\frac{2\pi}{a_M}, -\frac{2\pi}{\sqrt{3}a_M})$. We used $V = -14.88$ meV and $\varphi = -50.19°$.

In the calculation, for a given magnetic flux, we first solve $H_0$ and get degenerate Landau levels. Then we project $H_M$ term in the Landau level basis and solve the resulting Hamiltonian following Ref. 23.

We report our calculated results in Fig.S3. Here we include both the spin and the valley degree of freedom. The spin Zeeman coupling is ignored in the calculation. First, for $D < 0$, we find that the $n < 0$ side does not have clear Hofstadter butterfly pattern. This is because of the following reason: For ABC trilayer graphene, the zeroth Landau level is layer polarized. With $D < 0$, the electrons in the conduction band are mainly from the top layer and the holes in the valence band are mainly from the bottom layer. Since the hBN moiré superlattice potential applies only on the top graphene layer, only the conduction band can feel strong moiré superlattice potential. Similarly, for the $D > 0$ side, only the $n < 0$ side shows clear Hofstadter pattern. When magnetic field $B$ is large, we can focus purely on the zeroth Landau level. In the $n > 0$, $D < 0$ side, the strong moiré superlattice potential splits the zeroth Landau level to Hofstadter subbands. From our calculation, there is always a clear gap at $n = 2n_0$, which survives to very large magnetic field. Such a gap is indeed observed in the experiment and can be explained within single particle picture. In contrast, the gaps at fractional filling observed in the experiment do not exist in our single particle calculation and need correlation effects.

**Method references**

23. Zhang, Y.-H., Po H. C. & Senthil, T. Landau level degeneracy in twisted bilayer graphene: Role of symmetry breaking. *Phys. Rev. B* **100**, 125104 (2019).

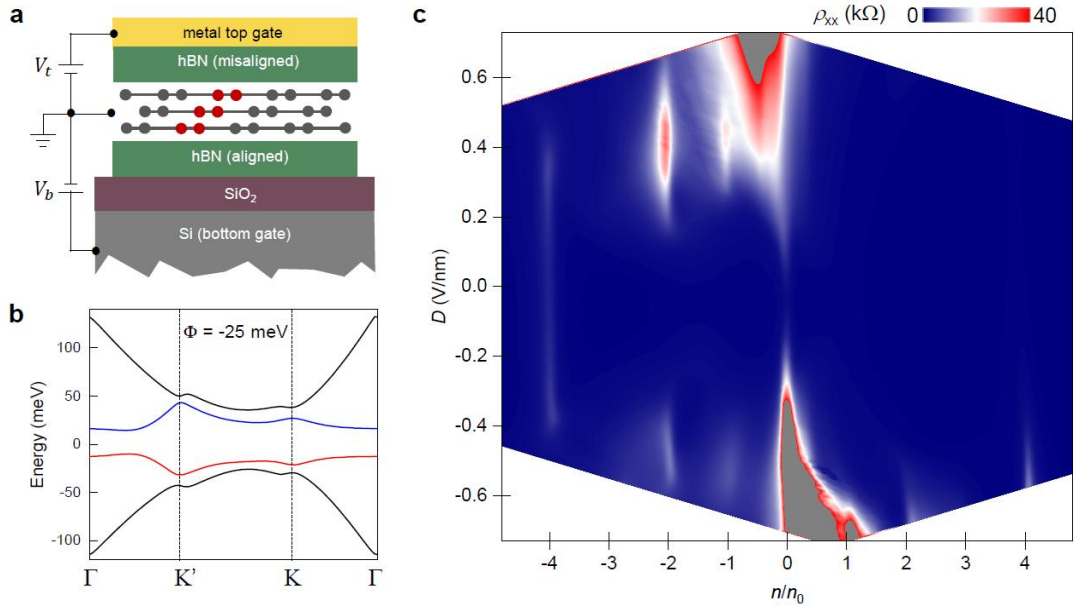

**Figure 1. ABC-TLG/hBN moiré heterostructure and its flat bands. a,** Schematic cross-sectional view of the device. The red atoms represent one unit cell of ABC-TLG. The ABC-TLG is aligned with the bottom hBN film. **b,** Calculated single-particle band structure of ABC-TLG/hBN in the K valley for a potential difference Φ = -25 meV between top and bottom graphene layer, which proximately corresponds to $D$ = -0.5 V/nm. **c,** Experimental phase diagram presented by the resistivity as a function of doping $n$ and vertical displacement field $D$ at $T$ = 1.5 K, $B_\perp$ = 0 T.

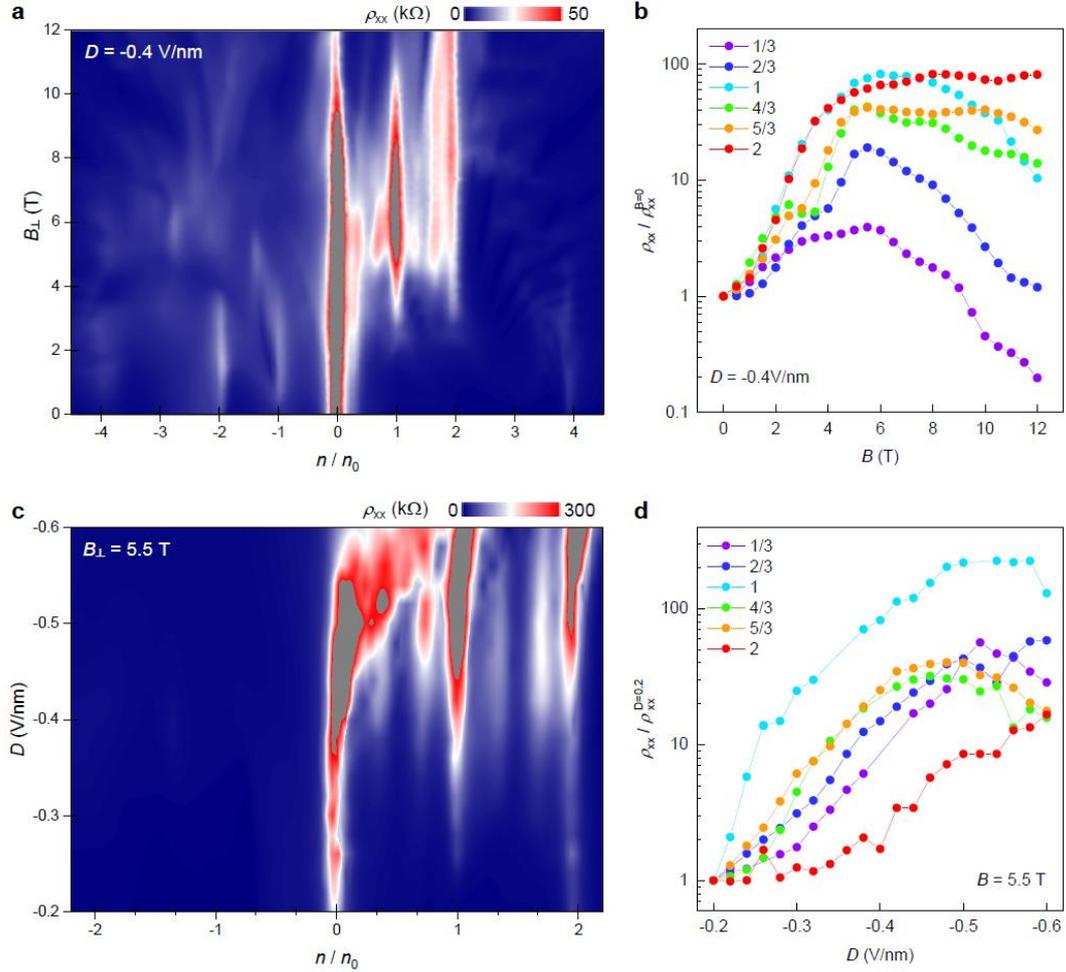

**Figure 2. The emergence of correlated states in the conduction minibands tuned by magnetic field and vertical displacement field. a,** Color plot of $\rho_{xx}$ as functions of $n/n_0$ and $B_\perp$ at the fixed $D$ = -0.4 V/nm. **b,** The normalized resistivity, $\rho_{xx} / \rho_{xx}^{B=0}$, as a function of the magnetic field at the fillings of $n_0/3$ to $2n_0$, where $\rho_{xx}^{B=0}$ is the resistivity at $B$ = 0 T. The data are extracted from **a**. **c,** Color plot of $\rho_{xx}$ as functions of $n/n_0$ and $D$ at the fixed $B_\perp$ = 5.5 T. **d,** $\rho_{xx} / \rho_{xx}^{B=0}$ as a function of the $D$ for the fillings of $n_0/3$ to $2n_0$. The data are extracted from **c**.

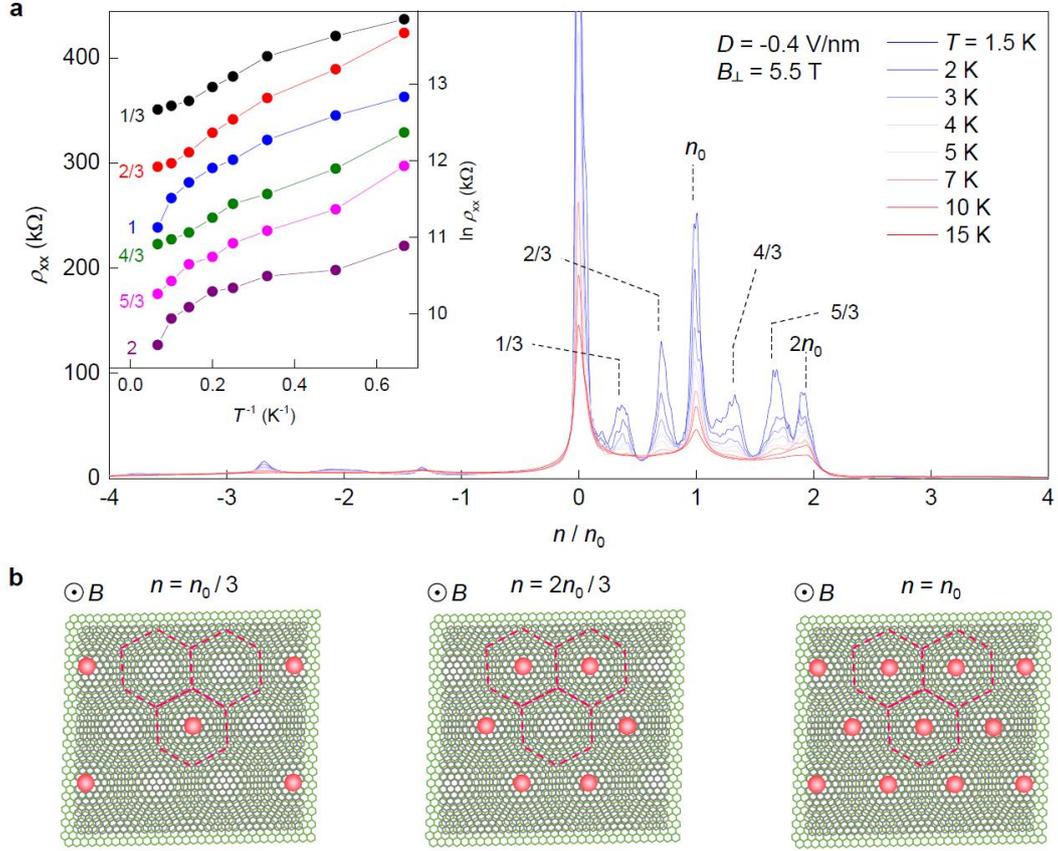

**Figure 3. Generalized Mott and Wigner crystal states at the conduction minibands in a magnetic field. a,** Resistivity, $\rho_{xx}$, as a function of the filling of the moiré site $n/n_0$ at the fixed $D = -0.4$ V/nm and $B_\perp = 5.5$ T at temperatures varying from 1.5 K to 15 K. Inset shows the thermal excitation behavior (Arrhenius plot) of the resistivity at the resistivity peaks at different fillings of the conduction minibands. **b,** Illustrations of the generalized Mott ($n = n_0$) and Wigner crystal ($n = n_0/3$, $n = 2n_0/3$) states in an out-of-plane magnetic field in real space. The grey and green lattices represent graphene and hBN lattices, red balls represent electrons, and the red dashed lines represent three moiré unit cells.

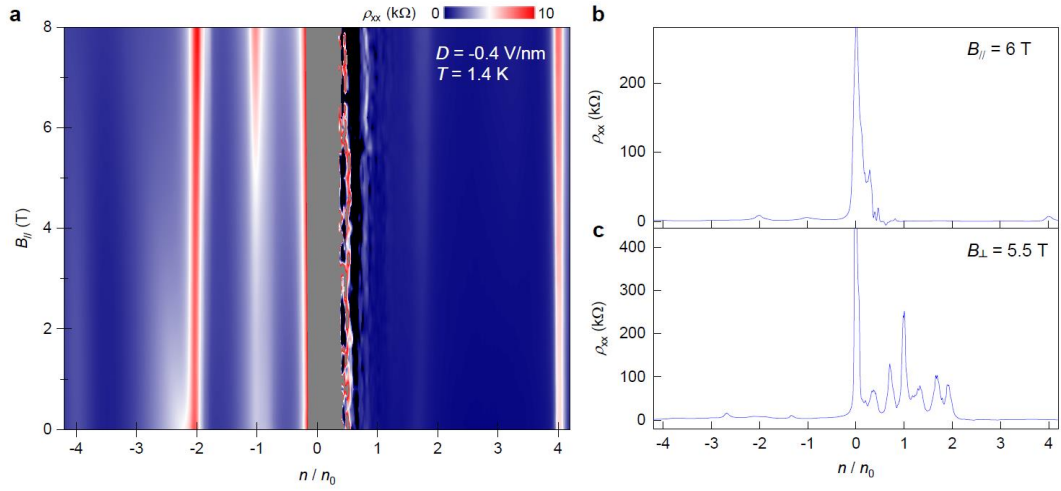

**Figure S1. Transport at the in-plane magnetic field. a,** Color plot of $\rho_{xx}$ as functions of $n/n_0$ and in-plane magnetic field $B_{//}$ at the fixed $D$ = -0.4 V/nm, $T$ = 1.5 K. **b,** $\rho_{xx}$ as a function of $n/n_0$ at $B_{//}$ = 6 T, which is extracted from the line cut of **a**. **c,** $\rho_{xx}$ as a function of $n/n_0$ at $B_\perp$ = 5.5 T, which is the same data with Fig. 3a.

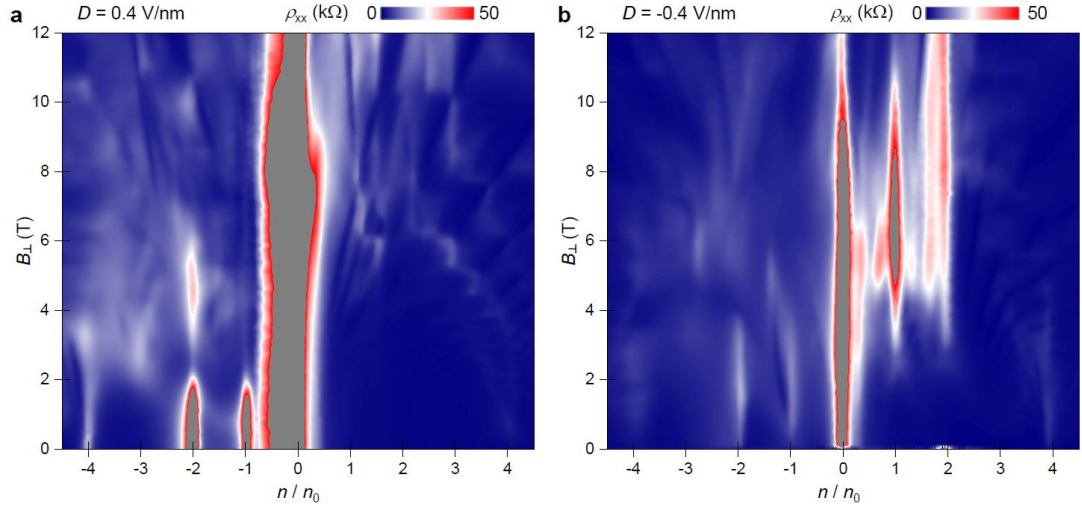

**Figure S2. Magneto transport at positive and negative *D*.** Color plot of $\rho_{xx}$ as functions of $n/n_0$ and vertical magnetic field $B_\perp$ at (**a**) *D* = 0.4 V/nm and (**b**) *D* = -0.4 V/nm, *T* = 1.5 K. The correlated insulating states at the conduction moiré bands only emerge at negative *D*, which is predicted to be topological nontrivial band.

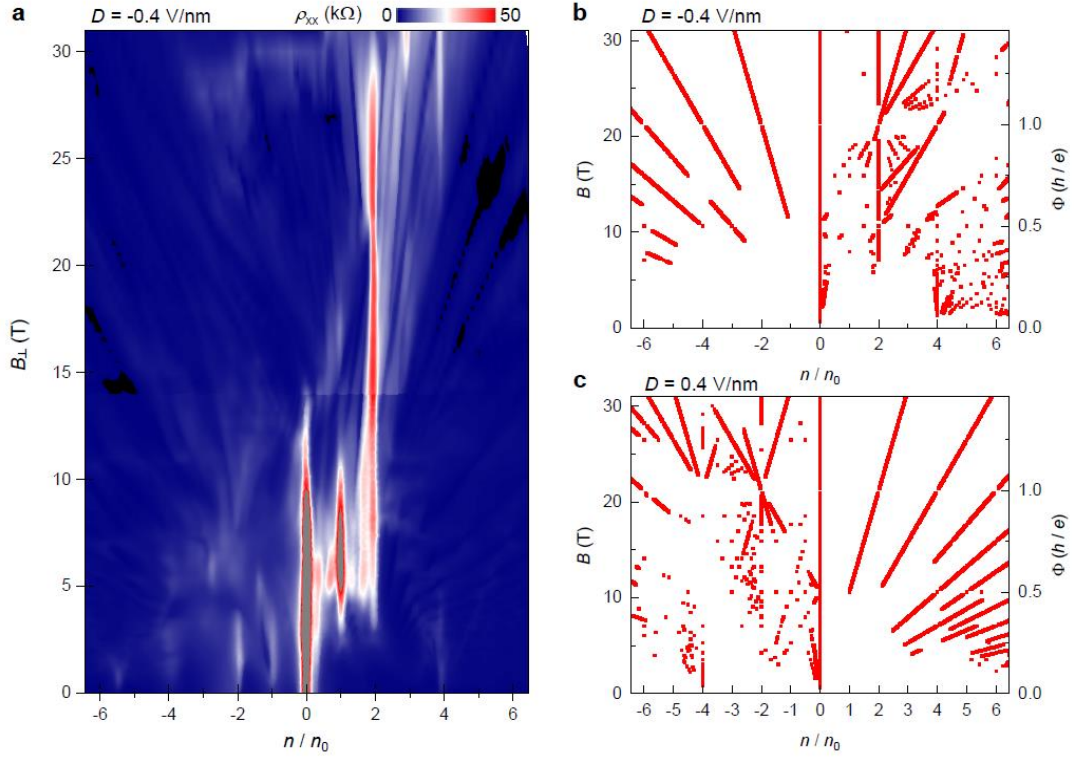

**Figure S3. Transport at high magnetic field. a,** Color plot of $\rho_{xx}$ as functions of $n/n_0$ and vertical magnetic field $B_\perp$ at the fixed $D = -0.4$ V/nm, $T = 1.5$ K for data below 14 T, and 0.3 K for data above 14 T. **b and c,** Calculated LL fan diagram for $D = -0.4$ V/nm and 0.4 V/nm, respectively. Data points represents the Hofstadter butterfly states with a gap larger than 3 meV.